\documentstyle[epsf]{elsart}
\begin{document}
\begin{frontmatter}
  \rightline{HUB-EP-99/23}\vskip 1truecm
  
  \title{Are there short distance non-perturbative contributions
  to the QCD static potential?}
  \author{Gunnar S.\ Bali\thanksref{email}}
  \thanks[email]{E-mail: bali@physik.hu-berlin.de}
  \address{Humboldt-Universit\"at zu Berlin, Institut f\"ur Physik,
Invalidenstr.~110, 10119 Berlin, Germany}

\begin{abstract}
\baselineskip=20pt
We find that perturbation theory fails to describe the short range static
potential as obtained from quenched lattice simulations, at least
for source separations $r>(6$~GeV)${}^{-1}$.
The difference between
the non-perturbatively determined potential and perturbation theory
at short distance is
well parameterised by a linear term with a slope of approximately
1.2~GeV${}^2$,
that is significantly bigger than the string tension, $\sigma\approx
0.21$~GeV${}^2$.
\end{abstract}
\end{frontmatter}


\section{Introduction}
The operator product expansion (OPE)
constitutes the standard framework of including non-perturbative
contributions to a QCD observable.
Within the OPE, non-perturbative
power corrections are included
in the form of infrared sensitive operators of a given dimension
that are accompanied by ultraviolet coefficient functions.
These Wilson coefficients are in general themselves only fixed up to
power corrections that can be reshuffled into higher dimensional
operators without affecting physical results. This ambiguity
is related to the presence of infrared renormalons~\cite{ope} and also
reflects the asymptotic character of the QCD perturbative series.

The powers of the correction terms within the OPE are
completely determined by the symmetries
of the operator that is expanded.
In a standard infrared renormalon analysis the dimension
of the lowest order
renormalon contribution
coincides with that of the lowest non-perturbative
condensate that contributes. In the case of the running coupling this
would for instance be the gluon condensate~\cite{sumrule}, such that
the leading order power correction to $\alpha_s(q)$ is expected to be
proportional to $\Lambda^4/q^4$.
This picture has recently been challenged by
the ultraviolet renormalon technique~\cite{uvren} and
dispersion approaches~\cite{zakharov,grunberg,shirkov}.
It has been argued that in addition to the
terms expected from the OPE, a further ultraviolet renormalon
contribution, proportional to
$\Lambda^2/q^2$ might be present.
Insofar, renormalons seem to allow 
for a more universal classification
of power-like corrections than the OPE~\cite{renorm}.
Power corrections to $\alpha_s$ have been known to
arise naturally in many physical
schemes for quite a while~\cite{brodsky}.

Evidence of an unexpected $\Lambda^2/q^2$ power correction
to the gluon condensate has been
obtained in the lattice study of
Ref.~\cite{burgio}. Recently, the running coupling from
the three-gluon vertex determined in Landau gauge on the lattice has been
investigated and also here evidence for a contribution proportional
to $\Lambda^2/q^2$ has been reported~\cite{burgio2}.

In case of the static potential in position space, such a 
$\Lambda^2/q^2$ contribution to the running coupling results in a term
proportional to the quark separation, $r$, while the leading order
correction from the standard renormalon analysis
depends only quadratically on $r$~\cite{zakharov3}. Indeed,
based on various reasonable model assumptions such linear
short-range contributions have
been obtained~\cite{zakharov3,simonov,zakharov4}.

The present letter is motivated by these latest developments 
and the phenomenological relevance
of the short-range static QCD potential in view of calculations
of the $t\overline{t}$ threshold cross-section.
We summarise recent
perturbative results, describe the method that we use to subtract
the QCD perturbative series and, indeed, find the non-perturbative
contribution to be dominated by a linear term.

\section{The potential in perturbation theory}
Recently, the static potential,
\begin{equation}
\label{Vq}
V(q)=-C_F\frac{4\pi\alpha_V(q)}{q^2},
\end{equation}
has been calculated to two loops by Peter~\cite{peter,peter2}.
This calculation has been independently
confirmed by Schr\"oder~\cite{york}
who corrected a normalisation error of the original reference.
The result reads,
\begin{equation}
\label{alphav}
\alpha_V(q)=\alpha_{\overline{MS}}(q)\left(1+
a_1\alpha_{\overline{MS}}(q) 
+a_2\alpha_{\overline{MS}}^2\right).
\end{equation}
For the quenched approximation ($n_f=0$) they obtain,
\begin{equation}
a_1=\frac{31}{12\pi},\quad
a_2=\left(\frac{4343}{18}+36\pi^2-\frac{9}{4}\pi^4+66\zeta(3)\right)
\frac{1}{16\pi^2}.
\end{equation}
$\zeta(3)\approx 1.202056903$ denotes the Riemann $\zeta$-function.
$a_1$ has already been calculated some time ago~\cite{fischler,billoire}.

A Fourier transformation yields,
\begin{equation}
\label{Vrun}
V(r)=-C_F\frac{\alpha_R(1/r)}{r},
\end{equation}
for the potential in position space with~\cite{billoire,peter2,peter3},
\begin{equation}
\label{alphaR}
\alpha_R(1/r)=\alpha_V(\mu)\left(1+\frac{\pi^2\beta_0^2}{3}\alpha_V^2\right),
\quad\mu=\exp(-\gamma_E)/r.
\end{equation}
$\gamma_E$ denotes the
Euler constant.
The coefficients of the QCD $\beta$-function for $n_f=0$ are given
by~\cite{tarasov},
\begin{equation}
\beta_0=\frac{11}{4\pi},\quad
\beta_1=\frac{102}{16\pi^2},\quad
\beta_2^{\overline{MS}}=\frac{2857}{128\pi^3}.
\end{equation}

By employing a recursive lattice finite size technique~\cite{fstech},
the ALPHA collaboration
has recently obtained a value for the running coupling in quenched
QCD~\cite{alpha},
\begin{equation}
\label{lambda}
\Lambda_{\overline{MS}}^{(0)}=0.602(48)\,r_0^{-1}.
\end{equation}
$r_0\approx 0.5$~fm denotes the Sommer~\cite{sommer} scale
of eq.~(\ref{som}) below.

We choose to parametrise
the solution of the renormalisation group
equation by,
\begin{equation}
\label{alpharun}
\alpha_{\overline{MS}}(\mu)=\frac{1}{\beta_0}\left[t
+d_1\ln t +d_1^2\frac{\ln t}{t}+\frac{d_2}{t}
-\frac{d_1^3}{2}\frac{\ln^2 t}{t^2}
+\frac{\beta_1\beta_2}{\beta_0^5}\frac{\ln t}{t^2}\right]^{-1},
\end{equation}
with $t=2\ln(\mu/\Lambda_{\overline{MS}})$,
$d_1=\frac{\beta_1}{\beta_0^2}$ and
$d_2=d_1^2-\frac{\beta_2}{\beta_0^3}$, which is
consistent to order $\alpha_s^3$.
The integration constant corresponds to the
standard normalisation convention of the QCD $\Lambda$-parameter.

\section{The lattice potential}
We have collected lattice data on the static potentials from
Refs.~\cite{balischilling,balischilling2,baliwachter,balitcl}.
All these results have been obtained in the quenched approximation
with Wilson action at values of the
bare lattice coupling $6.0\leq \beta=6/g_0^2\leq 6.8$ that correspond
to lattice spacings $0.19\,r_0>a>0.069\,r_0\approx 0.035$~fm. The linear
lattice extents are kept at a similar size in physical units and
range from $L=16$ at $\beta=6.0$ to $L=48$ at $\beta=6.8$.
In all cases, we find $La>3\,r_0\approx 1.5$~fm, such that finite
size effects are negligible~\cite{balischilling,balischilling0}.

All lattice data within this $\beta$-range
can be well fitted to the Cornell ansatz,
\begin{equation}
\label{fit}
aV(r)=aV_0(a)-\frac{e}{ra^{-1}}+(\sigma a^2) (ra^{-1})
\end{equation}
for $r>0.4\,r_0$, where rotational invariance is found to be
restored within
statistical errors. The parameter $V_0(a)$ that contains
the self energy of the two static sources diverges like
$-1/[a\ln(\Lambda a)]$ as the continuum limit is taken.
The scale $r_0$ is defined by~\cite{sommer},
\begin{equation}
\label{som}
\left.\frac{dV(r)}{dr}r^2\right|_{r=r_0} = 1.65.
\end{equation}
From quarkonium phenomenology one finds~\cite{sommer,baliwachter},
$r_0\approx 0.5$~fm. We obtain this scale from our fits
to eq.~(\ref{fit}):
$r_0a^{-1}=\sqrt{(1.65-e)/(\sigma a^2)}$. 

\begin{figure}[htb]
  \makebox[2.5truecm]{\phantom b} \epsfxsize=12truecm
  \epsffile{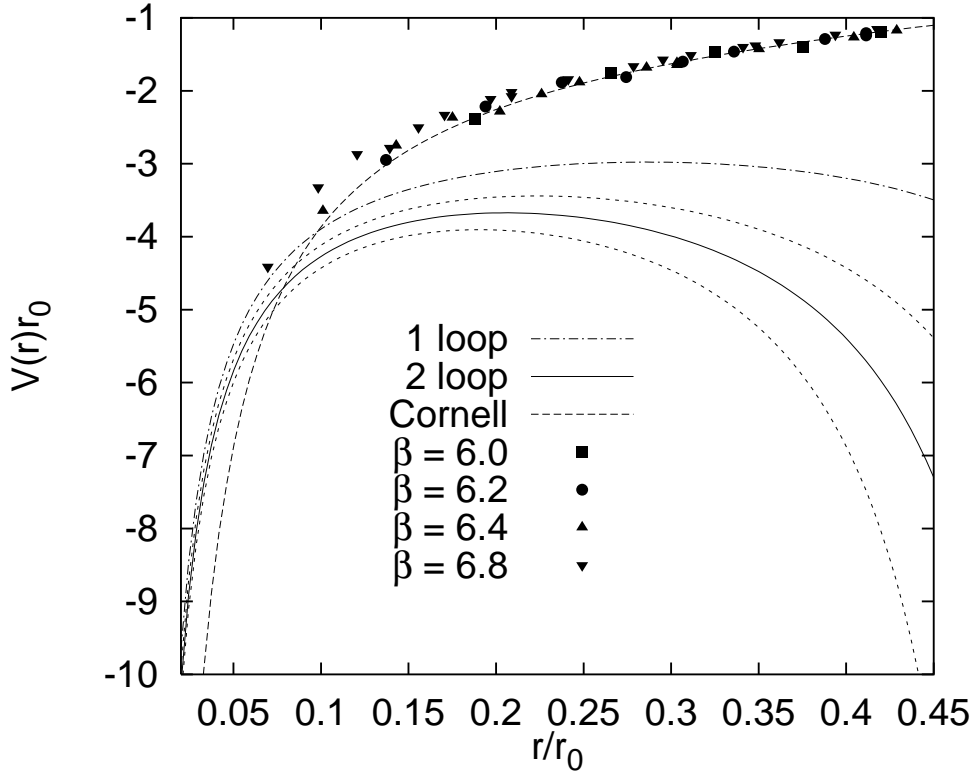}
\caption{ The quenched lattice potential versus
perturbation theory in units of $r_0\approx 0.5$~fm.
The only free parameter is the self-energy of the
static sources. The lattice data has been normalised such that $V(r_0)=0$.
The values $0.94\,r_0^{-1}$ and $0.77\,r_0^{-1}$
have been subtraced from the perturbative one- and two-loop
formulae, respectively, to allow for better comparison.}
\label{figpertpot}
\end{figure}

In Fig.~\ref{figpertpot}, we plot the short range
part of the lattice potentials obtained at different lattice
spacings $a_{\beta}$ in
units of $r_0$. The statistical errors are all smaller than the
corresponding data symbols.
The normalisation convention adopted to cancel the
self-energy is $V(r_0)=0$. The first four data points
measured at each $\beta$-value
correspond to $r/a_{\beta}=1,\sqrt{2},\sqrt{3},2$,
respectively.
The data for $r>2a_{\beta}$ all follow a universal
continuous curve
and rotational invariance is
restored qualitatively. We take this as indication that
for these distances, the continuum limit is effectively
realised.
For comparison, the result from the Cornell fit to $r>0.4\,r_0$,
used to determine the scale $r_0$,
is plotted (dashed curve). Towards small $r$ the data points lie
above this curve, indicating asymptotic freedom, i.e.\
a weakening of the effective Coulomb
coupling with the distance.

In addition, we have included the one-loop (dashed dotted curve) and
two-loop (solid curve with error band)
continuum expectation
of eqs.~(\ref{alphav})--(\ref{alpharun}).
Note that,
up to the additive self-energy constant, no free parameter can be
adjusted.
For better comparison with the lattice data,
normalised to $V(r_0)=0$, we have subtracted the values $0.94\,r_0^{-1}$
and $0.77\,r_0^{-1}$ from the one- and two-loop expectations, respectively.
The perturbative expression has a pole at
$\Lambda_{\overline{MS}}^{-1}\approx 1.66\,r_0\approx (240$~MeV$)^{-1}$,
such that we 
would not expect perturbation theory to reliably describe the lattice data
for $r>0.5\,r_0\approx 0.25$~fm or so. As can be seen from the plot,
even at the smallest
distance at our disposal,
$r\approx 0.07\,
r_0\approx(24\,\Lambda_{\overline{MS}})^{-1}$, the lattice
data substantially deviates from
perturbation theory; at any given distance the perturbative formula
considerably under-estimates the slope $dV/dr$.

In Ref.~\cite{peter4} it has been pointed out that due to the different
form of higher
order corrections, depending on whether one calculates the potential in
position or momentum space, numerically significant
differences arise. We have convinced ourselves that in the quenched theory
for $0.05\, r_0<r<0.4\, r_0$ this systematic
uncertainty can indeed be numerically up to two
times as big as the
error due to the uncertainty in $\Lambda_{\overline{MS}}$
displayed in the plot. However, we found that
the observed disagreement with perturbation
theory cannot be repaired by fiddling around with the
parametrisation. We remark that the difference between one- and
two-loop perturbation theory is quite big as indicated by the value,
$\beta_2^V=\beta_2^{\overline{MS}}+a_1\beta_1+(a_2-a_1^2)\beta_0\approx
3.19\gg \beta_2^{\overline{MS}}\approx 0.72$.

In position space, the correction is even more significant: starting
from eqs.~(\ref{alphav}) and (\ref{alphaR}) and the definition,
\begin{equation}
\alpha_R(1/r)=\alpha_{\overline{MS}}(1/r)\left[1+c_1\alpha_{\overline{MS}}(1/r)
+c_2\alpha^2_{\overline{MS}}\right],
\end{equation}
we obtain,
\begin{equation}
\label{error}
c_1=a_1+2\gamma_E\beta_0,\quad
c_2=a_2+4\gamma_E\beta_0a_1+2\gamma_E\beta_1+
\left(4\gamma_E^2+\frac{\pi^2}{3}\right)\beta_0^2.
\end{equation}
This results in the numerical value, $\beta_2^R=\beta_2^{\overline{MS}}
+c_1\beta_1+(c_2-c_1^2)\beta_0\approx 6.70$.

\section{Non-perturbative ultraviolet contributions}
Given the asymptotic character of QCD perturbative series, it is
always ambiguous how to subtract perturbative contributions to a
given physical process in order to isolate non-perturbative terms.
Bearing this in mind, we employ the procedure
of subtracting the one- and two-loop result, given by
eqs.~(\ref{alphav})--(\ref{alpharun}). Needless to say, that using a
different scheme or cutting off the series
at a different order in $\alpha_s$
will in general
yield somewhat different results.
We would like to remark that
beyond the two-loop level there appear to be difficulties in
a consistent perturbative definition of the static
potential~\cite{appelquist,soto}.

\begin{figure}[htb]
  \vskip 0.8truecm \makebox[2.5truecm]{\phantom b} \epsfxsize=12truecm
  \epsffile{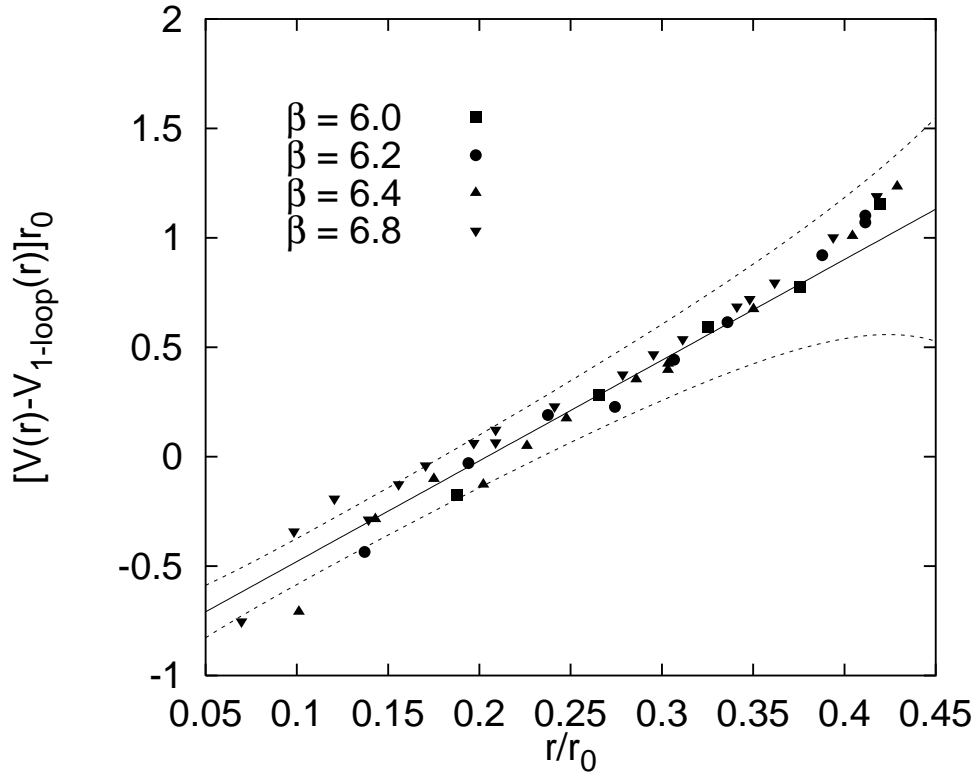}
\caption{The quenched lattice potential with one-loop
perturbation theory subtracted, in
comparison with a linear curve with slope $4.6\,r_0^{-2}$.}
\label{figpertpot3}
\end{figure}

\begin{figure}[htb]
  \vskip 0.8truecm \makebox[2.5truecm]{\phantom b} \epsfxsize=12truecm
  \epsffile{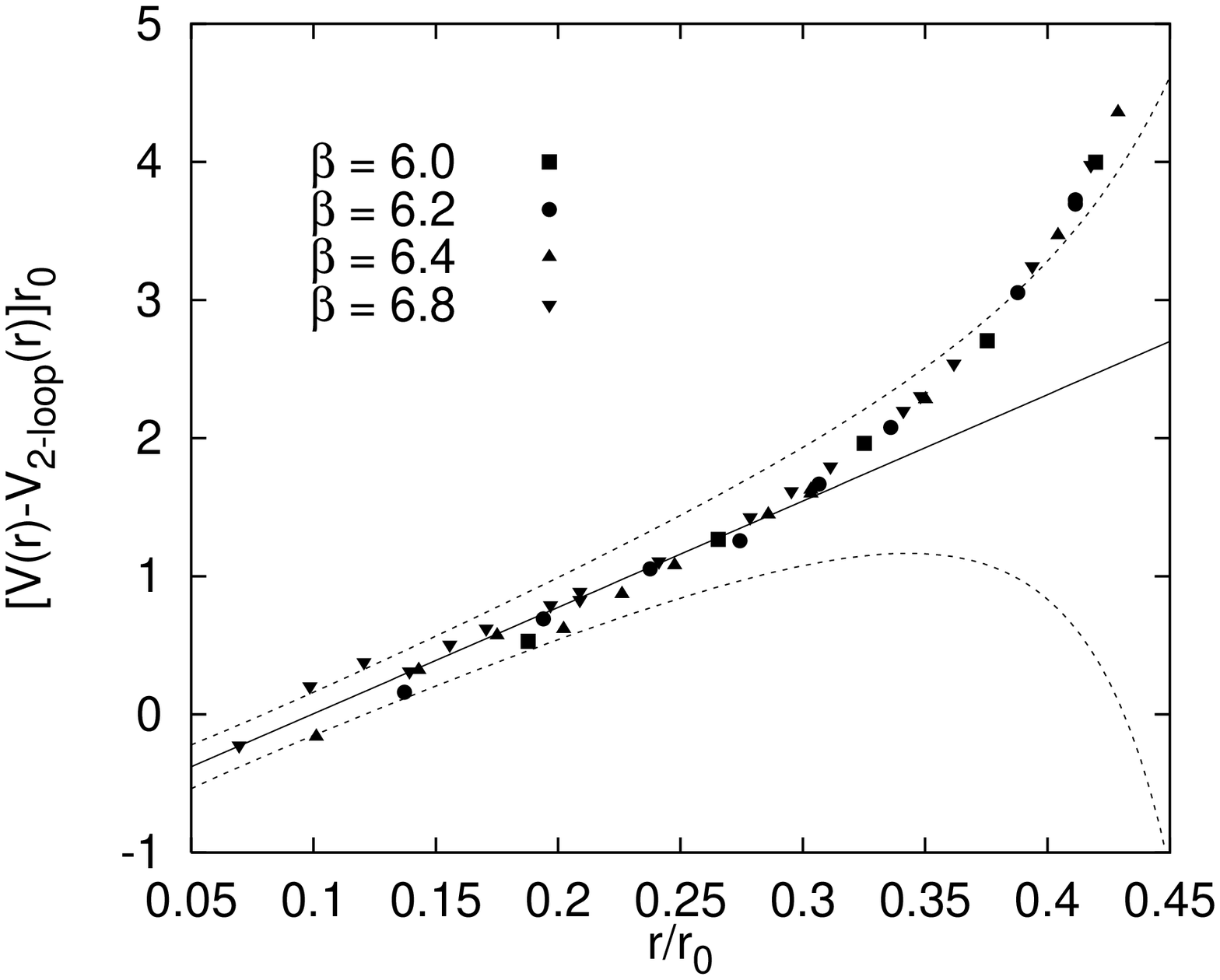}
\caption{The quenched lattice potential with two-loop
perturbation theory subtracted, in
comparison with a linear curve with slope $7.7\,r_0^{-2}$.}
\label{figpertpot2}
\end{figure}

In Figs.~\ref{figpertpot3} and
\ref{figpertpot2}, we have subtracted the one- and two-loop
continuum perturbation theory result that correspond to the
central value $\Lambda_{\overline{MS}}=0.602\,r_0^{-1}$
from our lattice data.
The resulting points are compared to linear curves (solid lines) with slopes
$4.6\,r_0^{-2}\approx 0.72$~GeV${}^2$ and $7.7\,r_0^{-2}\approx
1.20$~GeV${}^2$, respectively. We have added
the error band due to the uncertainty in $\Lambda_{\overline{MS}}$ to these
curves. For $r>0.4\,r_0$ and $r>0.3\,r_0$ for the one- and two-loop results,
respectively, deviations from the linear behaviour become visible while
for $r$ values that are small in units of the lattice spacings
$a_{\beta}$ the points scatter around the interpolating curve
due to lattice artefacts.
The data indicate the dominant short range correction to perturbation
theory to be proportional to $r$. However, additional quadratic
contributions might very well be present too.

Given the visible lattice structure at small $r$, we do not attempt
a fit to the data but
we conservatively estimate systematic uncertainties
of up to 30~\% on the observed slopes
by allowing the interpolating curves to touch the most extreme
combinations of data points that are possible. We checked that
this error comfortably accomodates possible effects of
additional quadratic contributions.
The results are $3.4\pm 1.0$ and
$5.7\pm 1.7$ times bigger than the string tension
$\sigma\approx 1.35\,r_0^{-2}$ for the one- and two-loop
analysis, respectively.
Neither a term of the size of the string tension
as suggested in Ref.~\cite{polikarpov} nor a
linear short-range term $\Delta V_2(r)=\frac{18}{2\pi}\alpha_s\sigma r$
predicted by Simonov~\cite{simonov} can alone account for this slope.
Thus, additional
non-perturbative linear contributions are likely to be present.

We would like to emphasize that the independent determination
of the QCD $\Lambda$-parameter by the ALPHA
collaboration~\cite{alpha} was essential
for our study: it is always possible
to effectively compensate part of the slope by increasing the
value of the
$\Lambda$-parameter that enters the perturbative formula.

We choose to parameterise the running coupling into,
\begin{equation}
\alpha_R(1/r)=\alpha_{R,2-loop}(1/r) + c_Rr^2
\end{equation}
and find $c_R=-(0.90\pm 0.27)$~GeV$^2$.
A transformation into momentum space yields,
$\alpha_V(q)=\alpha_{V,2-loop}(q) + c_V/q^2$
with $c_V=-c_R/2$.
It is not {\em a priori} clear how to convert the result into
different schemes and how to apply it to different physical
observables. In fact, even the conversion to momentum space is not
exact since the Fourier transformation affects
higher order contributions and thus the
difference between our result and perturbation theory.
This can then result in a somewhat different slope
of the linear term, due
to the asymptotic character of the perturbative series.

\section{Conclusion}
We find that perturbation theory fails to describe the short range static
potential as obtained from quenched lattice simulations, at least
for source separations bigger than (6~GeV)${}^{-1}$.
The difference between
the non-perturbatively determined potential and
perturbation theory
at short distance is
well parameterised by a linear term.
We quote the two-loop slope,
$(1.20\pm 0.36)$~GeV${}^2$ as our final result, which
is significantly bigger than the string tension $\sigma\approx
0.21$~GeV$^2$.

We remark that the slope of the linear
short distance contribution to the potential
is, due to the large size of the two-loop coefficient,
somewhat affected by the details
of the resummation of the perturbative series.
The precise value also depends
on whether one allows for possible
subleading non-perturbative contributions
that are quadratic in $r$. 
This latter source of
uncertainty is included into our error estimate.

It is certainly worthwhile to repeat the present calculation
using lattice actions different from the Wilson action
to confirm universality of the result.
A further significant reduction of the lattice spacing is
at present ruled out by the size of available computers as
is a similar study with inclusion of sea quarks. In order to
keep the argumentation transparent, we did not attempt to correct
the potential for perturbatively known tree-level or one-loop
lattice artefacts. This can in principle be done to
reduce the amount of violations of rotational invariance in the
short distance lattice potential.
Unfortunately, no two-loop lattice perturbation theory
result on the potential
exists so far, such that a direct study of
renormalon contributions on the lattice rather than in the continuum
is not yet consistently possible to order $\alpha_s^3$.

It should be interesting to investigate the same question in 
$SU(N)$ gauge theory in
three space-time dimensions (which is super-renormalisable)
as well as to study the behaviour on small lattice
volumes, where it is possible to further reduce the lattice spacing.
As a consequence of an altered vacuum structure,
confinement is lost on spatial volumes smaller than about 1~fm$^3$.
However, contributions that are related to ultraviolet physics
should still be present and could
easily be disentangled from terms that arise due to
infrared properties of the theory.

\section*{Acknowledgments} 
G.B. acknowledges motivation by M.I.~Polikarpov and V.I.~Zakharov
who introduced him to the problem and inspiration from
discussions with N.~Brambilla, J.~Soto and A.~Vairo.
G.B. expresses thanks to A.~Pineda, N.~Brambilla and A.~Vairo
for detecting a sign error in eq.~(\ref{error}) of an earlier version of
this letter.
G.B. has been supported by DFG grant Nos.\ Ba 1564/3-1 and Ba 1564/3-2.


\begin{thebibliography}{999}
\bibitem{ope} G.~`t Hooft, in: The Whys of Subnuclear Physics,
Proc.\ Int.\ School, Erice 1977, ed.\ A.\ Zichichi,
(Plenum, New York, 1977); B.\ Lautrup, Phys.\ Lett.\ 69B, 109 (1977);
A.H.~Mueller, Nucl.\ Phys.\ B 250, 327 (1985);
A.H.~Mueller, in: QCD: 20 years later, ed.\ P.M.\ Zerwas and H.M.\ Kastrup,
(World Scientific,
Singapore, 1993); M.~Beneke and V.I.~Zakharov, Phys.~Lett.~B 312, 340 (1993).
\bibitem{sumrule} M.A.\ Shifman, A.I.\ Vainshtein, V.I.\ Zakharov,
Nucl.\ Phys.\ B 147, 385 (1978).
\bibitem{uvren} V.I.\ Zakharov, Nucl.\ Phys.\ B 385, 385 (1992);
A.I.\ Vainshtein and V.I.\ Zakharov, Phys.\ Rev.\ Lett.\ 73, 1207 (1994).
\bibitem{zakharov} R.\ Akhoury and V.I.\ Zakharov, in: Proc.\
5th International Conference on Physics beyond the
Standard Model, Balholm 1997,
ed.\ G.\ Eigen, P.\ Osland and B.\ Stugu, (AIP, Woodbury, 1997),
available as hep-ph/9705318;
Nucl.\ Phys.\ B (Proc.\ Suppl.) 64, 350 (1998). 
\bibitem{grunberg} G.\ Grunberg, hep-ph/9705290; 
in Proc.\
32nd Rencontres de Morino: QCD and High-Energy Hadronic Interactions,
Les Arcs 1997, ed.\ J.\ Tran Thanh Van, (Ed.\ Frontieres, Paris,
1997), available as hep-ph/9705460;
JHEP 9811, 6 (1998).
\bibitem{shirkov} D.V.\ Shirkov and I.L.\ Solovtsev, Phys.\ Rev.\ Lett.\
79, 1209 (1997);
D.V.\ Shirkov, Nucl.\ Phys.\ B (Proc.\ Suppl.) 64, 106 (1998).
\bibitem{renorm} See e.g., V.I.\ Zakharov,
Prog.\ Theor.\ Phys.\ Suppl.\ 131, 107 (1998).
\bibitem{brodsky} S.J.\ Brodsky, G.P.\ Lepage, P.B.\ Mackenzie,
Phys.\ Rev.\ D 28, 228 (1983).
\bibitem{burgio} G.\ Burgio, F.\ Di Renzo, G.\ Marchesini, E.\ Onofri,
Phys.\ Lett.\ B 422, 219 (1998).
\bibitem{burgio2} G.\ Burgio, F.\ Di Renzo, C.\ Parrinello, C.\ Pittori,
hep-ph/9808258.
\bibitem{zakharov3} R.\ Akhoury and V.I.\ Zakharov,
Phys.\ Lett.\ B 438, 165 (1998).
\bibitem{simonov} Yu.A.\ Simonov, hep-ph/9902233.
\bibitem{zakharov4} K.G.\ Chetyrkin, S.\ Narison, V.I.\ Zakharov,
Nucl.\ Phys.\ B 550, 353 (1999).
\bibitem{peter}
M.\ Peter, Phys.\ Rev.\ Lett.\ 78, 602 (1997).
\bibitem{peter2} M.\ Peter, Nucl.\ Phys.\ B 501, 471 (1997).
\bibitem{york} Y.\ Schr\"oder, 
Phys.\ Lett.\ B 447, 321 (1999).
\bibitem{fischler} W.\ Fischler, Nucl.\ Phys.\ B 129, 157 (1977).
\bibitem{billoire} A.\ Billoire, Phys.\ Lett.\ 92B, 343 (1980).
\bibitem{peter3} M.\ Je\.zabek, M.\ Peter, Y.\ Sumino,
Phys.\ Lett.\ B 428, 352 (1998).
\bibitem{tarasov} O.V.\ Tarasov, A.A.\ Vladimirov, A.Yu.\ Zharkov, Phys.\
 Lett.\ 93B, 429 (1980);
S.A.\ Larin, J.A.M.\ Vermaseren, Phys.\ Lett.\ B 303, 334
(1993).
\bibitem{fstech} M.\ L\"uscher, R.\ Narayanan, P.\ Weisz, U.\ Wolff, Nucl.\
 Phys.\ B 384, 168 (1992); M.\ L\"uscher, R.\ Sommer, P.\ Weisz, U.\ Wolff,
Nucl.\ Phys.\ B 491, 323 (1997).
\bibitem{alpha} ALPHA collaboration:
S.\ Capitani, M.\ L\"uscher,
R.\ Sommer, H.\ Wittig, 
Nucl.\ Phys.\ B 544, 669 (1999).
\bibitem{sommer}
R.\ Sommer, Nucl.\ Phys.\ B 411, 839 (1994).
\bibitem{balischilling} G.S.~Bali and K.~Schilling, Phys.\ Rev.\ D 47, 661
(1993).
\bibitem{balischilling2} G.S.~Bali and K.~Schilling, Int.\ Journ.\ Mod.\
Phys.\ C 4, 1167 (1993).
\bibitem{baliwachter} G.S.\ Bali, K.\ Schilling, A.\ Wachter,
Phys.\ Rev.\ D 56, 2566 (1997)
\bibitem{balitcl} $T\chi L$ collaboration: G.S.\ Bali et al.,
Nucl.\ Phys.\ B (Proc.\ Suppl.) 63, 946 (1998). 
\bibitem{balischilling0} G.S.~Bali and K.~Schilling, Phys.\ Rev.\ D 46, 2636
(1992).
\bibitem{peter4} M.\ Je\.zabek, J.H.\ K\"uhn, M.\ Peter, Y.\ Sumino, T.\
 Teubner, Phys.\ Rev.\ D 58, 14006 (1998).
\bibitem{appelquist} T.\ Appelquist, M.\ Dine, I.J.\ Muzinich,
Phys.\ Rev.\ D 17, 2074 (1978).
\bibitem{soto} N.\ Brambilla, A.\ Pineda, J.\ Soto, A.\ Vairo,
hep-ph/9903355.
\bibitem{polikarpov} F.V.\ Gubarev, M.I.\ Polikarpov, V.I.\ Zakharov,
hep-th/9812030.
\end{thebibliography}
\end{document}